\documentclass{ws-ijmpc}
\usepackage[super]{cite}
\usepackage{xcolor}
\usepackage{lineno}
\usepackage{booktabs}
\usepackage{multicol, multirow}
\usepackage[verbose,hypertexnames=false]{hyperref}
\hypersetup{colorlinks=false,allbordercolors=black,pdfborderstyle={/S/U/W 1}}

\begin{document}

\markboth{R.~Muslim}
{Topology-dependent criticality in triplet majority-rule dynamics with collective reversal}


\title{Topology-dependent criticality in triplet majority-rule dynamics with collective reversal on quenched networks}

\author{Roni Muslim}
\address{Asia Pacific Center for Theoretical Physics\\
Pohang 37673, Republic of Korea \\ Research Center for Quantum Physics,  BRIN,\\
South Tangerang, 15314, Indonesia \\
roni.muslim@apctp.org}

\maketitle

\begin{abstract}
We study a triplet majority-rule opinion-dynamics model with collective reversal
on quenched networks. Interactions occur on local triplets composed of one agent
and two of its neighbors, while collective reversal acts only on unanimous
triplets. This rule separates local conformity from external perturbations that
disrupt local agreement. We show that quenched network topology shifts the
order--disorder critical point away from the well-mixed value.
\textcolor{black}{For Barabási--Albert, Erdős--Rényi, random regular, and
Watts--Strogatz networks, the estimated critical exponents remain close to the
mean-field values, suggesting mean-field-like universal behavior within the
system sizes studied.}
\textcolor{black}{The strongest shift of the critical point occurs for
Watts--Strogatz networks, where clustering and local correlations make the
ordered phase less stable.}
A rewiring analysis of Watts--Strogatz networks further shows that the ordered
phase becomes more stable as the network becomes more random.
\textcolor{black}{These results indicate that quenched topology primarily controls
the location of the transition, while the collective-reversal mechanism largely
preserves mean-field-like critical behavior.}
\end{abstract}


\ccode{PACS Nos.: 89.65.-s, 89.75.Hc, 05.70.Fh, 05.10.Ln, 64.60.aq
}

\section{Introduction}
\label{sec:introduction}

Opinion dynamics is an important topic in sociophysics because it provides a
minimal framework for understanding how local interactions among individuals can
generate collective behavior at the population level. In agent-based approaches,
individuals are usually represented by discrete opinion variables, while opinion
changes are governed by microscopic update rules. Despite their simplicity, such
models can reproduce various macroscopic phenomena, such as consensus,
coexistence, polarization, and order--disorder transitions
\cite{clifford1973model,holley1975ergodic,galam2008sociophysics,
castellano2009statistical,sirbu2016opinion,noorazar2020classical,
noorazar2020recent,peralta2025opinion,flache2017models}.
Binary-opinion models, with $s_i=+1$ or $s_i=-1$, provide a useful starting point
for describing competition between two opposing social choices, such as support
and opposition, adoption and resistance, or trust and distrust.

One of the basic mechanisms in opinion formation is local conformity. In
majority-rule dynamics, a small group of agents interacts and the majority
opinion within the group determines the direction of the update. This mechanism
differs from the voter model, which is based on imitation of a single neighbor,
because the update is determined by the opinion composition of a local group
\cite{galam2002minority,krapivsky2003dynamics,chen2005majority,
castellano2009nonlinear,nyczka2012phase,radosz2017q}.
Thus, majority-rule dynamics can be viewed as a minimal representation of small
group discussions, local social pressure, or the tendency of individuals to follow
the dominant view in their social environment.

{\color{black}
Recent extensions of majority-rule and related opinion-dynamics models have
introduced several distinct mechanisms. Directed propaganda imposes an externally
selected opinion on an interacting group, independently of its initial
composition~\cite{forgerini2024directed}. Ratio-dependent contrarian activation
makes the response against the majority depend on whether the local group is
unanimous~\cite{galam2026ratio}. In structured Ising
social networks, nonfixed agents respond to the weighted majority of their
neighbors subject to a conviction threshold, while influential nodes retain fixed
opinions~\cite{bukina2026opinion}. These studies show that both the microscopic
response mechanism and the interaction structure can affect collective ordering
and phase-transition behavior.
}

Most analytical formulations of majority-rule and related opinion models have
been developed in the well-mixed or annealed setting, where interaction groups
are randomly selected from the entire population and their statistics depend only
on the global opinion density. Real social interactions, however, generally occur
through relatively persistent contact structures, commonly represented by
scale-free and small-world networks
\cite{barabasi1999emergence,watts1998collective}. Studies on random and
heterogeneous networks have shown that connectivity and topology can shift the
transition point and modify collective ordering
\cite{pereira2005majority,lima2007majority,alencar2024critical,
mulya2024phase}. Pair-approximation analyses further highlight the effects of
degree heterogeneity, clustering, and local correlations beyond homogeneous
mean-field descriptions
\cite{jkedrzejewski2017pair,jkedrzejewski2022pair}. Quenched structures are
also relevant to digital echo chambers, where persistent links contribute to
information segregation and locally homogeneous opinion groups
\cite{cinelli2021echo}.

{\color{black}
In addition to local-neighbor influence, opinion dynamics may include different
forms of nonconformity and external perturbation. Under independence, an agent
disregards the local configuration and updates autonomously or randomly
\cite{nyczka2012phase,abramiuk2020generalized, muslim2024impact,anugraha2025nonlinear}. Anticonformity and contrarian
behavior instead depend on social information, with an agent rejecting the local
majority and adopting the opposite or minority opinion
\cite{nyczka2013anticonformity,muslim2024impact}. Social noise represents random
errors or spontaneous opinion changes
\cite{mullick2025social}, whereas external fields introduce a
directional bias toward a prescribed opinion, often interpreted as mass-media
influence
\cite{civitarese2021external,azhari2023external,muslim2024mass}. In contrast,
the collective-reversal mechanism considered here acts only on unanimous
triplets and reverses all three opinions simultaneously, while mixed triplets
always follow majority rule. It is therefore a configuration-selective,
group-level perturbation that reorients a unanimous group without destroying its
internal agreement; in the symmetric case, it also favors neither opinion.
}

Specifically, we study triplet majority-rule dynamics on quenched networks. In
each elementary update, a central node $r$ is selected at random, and two of its
neighbors $j,\ell\in\partial r$ are selected to form a local triplet
$\tau=(r,j,\ell)$. Mixed triplets are updated according to the majority rule,
whereas unanimous triplets may undergo collective reversal. The formulation on
quenched networks introduces challenges that do not arise in well-mixed
populations. In the well-mixed case, the probability of obtaining a triplet with a
given number of positive spins follows a binomial form that depends only on the
global density $c$. In contrast, on quenched networks, two configurations with
the same value of $c$ may have different local arrangements. As a result, triplet
frequencies are no longer determined by $c$ alone, but also by the network
structure and the local correlations generated during the dynamics. This makes a
homogeneous mean-field approximation insufficient for capturing the influence of
topology on the critical point and collective behavior.

To address this issue, we construct a projected master-equation description in
the aggregate variable $q$, defined as the number of agents holding opinion $+1$.
Network information is retained through the conditional local-triplet statistics
$R_a(q;\epsilon)$, where $a=0,1,2,3$ denotes the number of positive spins in the
selected triplet. In contrast to the binomial mean-field approximation, the
transition rates in this description are determined by the actual triplet
statistics on quenched networks. Methodologically, $R_a(q;\epsilon)$ is
interpreted as a local statistic obtained from stationary microscopic
configurations, rather than as a direct fit to the order-parameter curve. The
approach is validated by comparing Monte Carlo (MC) simulations, the full master
equation for small systems, and the projected master equation.

The main contributions of this article are threefold. First, we study a triplet
majority-rule model on quenched networks with collective reversal on unanimous
triplets. Second, we show that the projected master equation based on
$R_a(q;\epsilon)$ can reproduce the aggregate dynamics obtained from MC
simulations and, for small systems, from the full master equation. Third, we
analyze the order--disorder transition on several network topologies, namely
Barabási--Albert (BA), Erdős--Rényi (ER), random regular (RR), and
Watts--Strogatz (WS) networks, using finite-size scaling
\cite{binder1981finite,privman1990finite,landau2021guide,
zubillaga2022three,mulya2024phase,alencar2024critical}.
{\color{black}
Our results show that topology shifts the critical point away from the well-mixed
value. For all network classes considered, the estimated critical exponents
remain close to the mean-field values, suggesting that collective reversal
primarily shifts the transition point while preserving mean-field-like critical
behavior within the system sizes studied. Among these networks, the WS topology
exhibits the strongest downward shift of the critical point, which is associated
with its persistent clustering and local correlations. This interpretation is
further supported by the rewiring analysis, which shows that the ordered phase
becomes more stable as the WS network becomes more random. Thus, the present
results indicate that quenched topology mainly controls the location of the
transition, rather than producing clear evidence of a distinct asymptotic
universality class.
}

This study is also related to recent developments in opinion dynamics based on
group interactions and higher-order interactions. Although the interaction
substrate in our model is a pairwise network, the update rule acts collectively
on local triplets. In this sense, the present model captures a higher-order-like
update mechanism without introducing explicit hyperedges
\cite{schawe2022higher,papanikolaou2023fragmentation,
iacopini2024temporal,kim2025competition,muslim2026effect,
battiston2020networks}.
Thus, the model provides a minimal framework for connecting local update rules,
triplet statistics on fixed networks, and macroscopic critical behavior.

\section{Model Description}
\label{sec:model}

We consider an opinion-dynamics model based on majority rule on a quenched
network. The system consists of $N$ agents located on the nodes of an undirected
network with adjacency matrix $A_{ij}$. Each agent $i$ carries a binary opinion
$s_i=\pm1$, where $s_i=+1$ and $s_i=-1$ represent two competing opinions. The
network structure is kept fixed throughout the dynamics, so interactions do not
occur randomly over the whole population, but are constrained by the existing
neighbor relations.

The degree of node $i$ is defined as $k_i=\sum_{j=1}^{N}A_{ij}$, and its set of
neighbors is denoted by $\partial i$. At each elementary update, a central node
$r$ is selected at random. Since the update requires two neighbors, only nodes
with $k_r\geq2$ can serve as the center of a triplet. If the selected node has
$k_r<2$, the central node is redrawn until an eligible node is obtained. Once the
central node $r$ is chosen, two distinct neighbors $j,\ell\in\partial r$ are
selected uniformly at random without replacement. Thus, the interaction group
always has size three, $n=3$, and the updated local triplet is
$\tau=(r,j,\ell)$.

The update rule consists of two mechanisms. The first mechanism is local
conformity through majority rule. If two of the three agents in the triplet hold
opinion $+1$, then all members of the triplet adopt opinion $+1$, namely
$++-\to +++$. Conversely, if two of the three agents hold opinion $-1$, then all
members of the triplet adopt opinion $-1$, namely $+--\to ---$. Since the group
size is always odd, no tie occurs in the majority update.

The second mechanism is collective reversal in the fully unanimous state. In a
unanimous state, majority rule produces no change because all members of the
triplet already hold the same opinion. However, an external influence can reverse
the opinion of the entire triplet collectively. Specifically, the transition
$---\to +++$ occurs with probability $\epsilon_{\uparrow}$, whereas the
transition $+++\to ---$ occurs with probability $\epsilon_{\downarrow}$, with
$0\leq\epsilon_{\uparrow},\epsilon_{\downarrow}\leq1$. If no reversal occurs,
the triplet remains in its original unanimous state. The parameter
$\epsilon_{\uparrow}$ represents an external tendency that drives opinion $-1$
toward $+1$, while $\epsilon_{\downarrow}$ represents an external tendency that
drives opinion $+1$ toward $-1$.

\begin{figure}[t!]
    \centering
    \includegraphics[width=\linewidth]{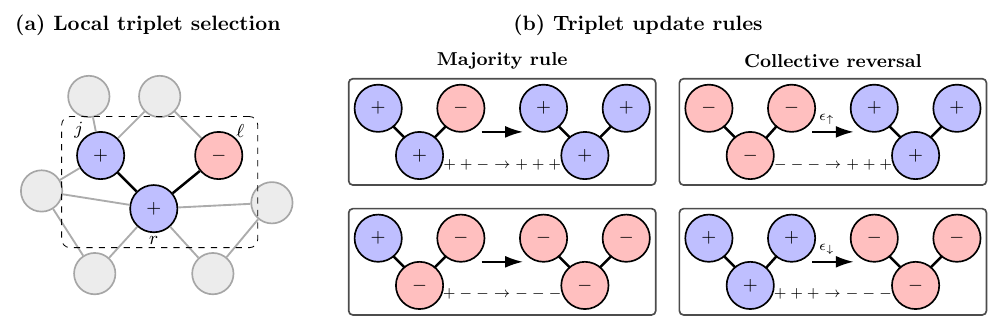}
    \caption{
    Illustration of the local triplet update on a quenched network.
    (a) A central node $r$ is selected at random from nodes that can form a
    triplet, and two distinct neighbors $j$ and $\ell$ are then selected to form
    the triplet $\tau=(r,j,\ell)$.
    (b) The triplet is updated either by majority rule, $++-\to +++$ and
    $+--\to ---$, or by collective reversal in a unanimous state,
    $---\to +++$ with probability $\epsilon_{\uparrow}$ and
    $+++\to ---$ with probability $\epsilon_{\downarrow}$.
    }
    \label{fig:illustration}
\end{figure}

In the symmetric collective-reversal case,
$\epsilon_{\uparrow}=\epsilon_{\downarrow}=\epsilon$, the two opinions are
subject to the same level of external perturbation. In contrast, when
$\epsilon_{\uparrow}\neq\epsilon_{\downarrow}$, the external perturbation is
asymmetric and favors one of the two opinions. The local triplet-selection and
update mechanisms are illustrated in Fig.~\ref{fig:illustration}.

This model differs from well-mixed or annealed formulations. In the well-mixed
setting, interaction groups are randomly selected from the whole population at
each update, so each agent can effectively interact with any other agent. In
contrast, on a quenched network, interactions are constrained by fixed local
connections. Therefore, the opinion dynamics is determined not only by the global
opinion fraction, but also by the network topology, degree heterogeneity, hubs,
clustering, and local correlations generated during the update process.

From a social perspective, a local triplet can be interpreted as a small
interaction unit, such as a discussion between an individual and two of their
nearest social contacts. Majority rule represents the tendency of individuals to
adjust their opinions to the local pressure of their small group. Collective
reversal in the unanimous state represents an external perturbation or
information source outside the local neighborhood, such as media, campaigns, or
macroscopic social pressure, that can change the opinion of a group even after it
has reached local agreement. Thus, the model captures the competition between
local conformity on a fixed social network and external perturbations that can
disrupt local consensus, a dynamic that has been observed empirically across various social media platforms~\cite{cinelli2021echo}.

\section{Results and Discussion}
\label{sec:results}

Before analyzing the critical behavior on large networks, we first construct an
aggregate description based on the master equation. The purpose is to validate
that the macroscopic dynamics observed in MC simulations can be
reproduced by a reduced master-equation description that preserves local triplet
statistics on quenched networks. After this validation, we analyze the order
parameter, susceptibility, Binder cumulant, and finite-size scaling behavior for
different network topologies.

\subsection{Master-equation formulation and aggregate description}
\label{subsec:master_equation}

The stochastic dynamics of the model can be described by a master equation in
the full configuration space. Let
$\mathbf{s}=(s_1,s_2,\ldots,s_N)$ denote a microscopic configuration of the
system, with $s_i=\pm1$, and let $P(\mathbf{s},t)$ be the probability that the
system is in configuration $\mathbf{s}$ at time $t$. The full master equation is
given by
\begin{equation}
\frac{dP(\mathbf{s},t)}{dt}
=
\sum_{\mathbf{s}'}
\left[
W(\mathbf{s}'\to \mathbf{s})P(\mathbf{s}',t)
-
W(\mathbf{s}\to \mathbf{s}')P(\mathbf{s},t)
\right],
\label{eq:full_master_equation}
\end{equation}
where $W(\mathbf{s}\to\mathbf{s}')$ denotes the transition rate from
configuration $\mathbf{s}$ to configuration $\mathbf{s}'$. This equation
expresses the balance between the probability current entering configuration
$\mathbf{s}$ and the probability current leaving it. Since transitions occur
only through local triplet updates, the explicit form of
$W(\mathbf{s}\to\mathbf{s}')$ is fully determined by the microscopic rule defined
in Sec.~\ref{sec:model}.

For a local triplet $\tau=(r,j,\ell)$, node $r$ is the central node, while
$j,\ell\in\partial r$ are two randomly selected neighbors. Since only nodes with
$k_r\geq2$ can form a triplet, we define
$\mathcal{V}_2=\{r\,|\,k_r\geq2\}$ and $N_2=|\mathcal{V}_2|$. In the
simulations, if the selected central node has $k_r<2$, the selection is repeated
until a node in $\mathcal{V}_2$ is obtained. Therefore, the selection weight of a
triplet is $\omega_{\tau}=1/[N_2\binom{k_r}{2}]$ for $r\in\mathcal{V}_2$.

The number of positive spins in the triplet $\tau=(r,j,\ell)$ is written as
$a_{rj\ell}(\mathbf{s})=(1+s_r)/2+(1+s_j)/2+(1+s_\ell)/2$, with
$a_{rj\ell}=0,1,2,3$. The values $a=0,1,2,3$ correspond respectively to the
triplet configurations $---$, $+--$, $++-$, and $+++$. The weighted fraction of
local triplets with $a$ positive spins in configuration $\mathbf{s}$ is then
given by
\begin{equation}
\Omega_a(\mathbf{s})
=
\frac{1}{N_2}
\sum_{r\in\mathcal{V}_2}
\frac{1}{\binom{k_r}{2}}
\sum_{\substack{j,\ell\in\partial r\\ j<\ell}}
\mathbf{1}
\left[
a_{rj\ell}(\mathbf{s})=a
\right],
\qquad
a=0,1,2,3.
\label{eq:Omega_a}
\end{equation}
With this definition, $\sum_{a=0}^{3}\Omega_a(\mathbf{s})=1$. The quantity
$\Omega_a(\mathbf{s})$ can be interpreted as the probability that a local triplet
selected according to the dynamical rule contains $a$ positive spins. Thus,
$\Omega_a(\mathbf{s})$ connects the microscopic configuration $\mathbf{s}$ to
the actual frequency of local triplets on the quenched network. For networks in
which all nodes satisfy $k_r\geq2$, such as the BA, RR, and WS networks used
here, one has $N_2=N$.

Let $q(\mathbf{s})=\sum_i(1+s_i)/2$ denote the number of positive spins. In one
elementary update, $q$ can change through four channels: $---\to +++$ gives
$\Delta q=+3$ with probability $\epsilon_{\uparrow}$, $++-\to +++$ gives
$\Delta q=+1$, $+--\to ---$ gives $\Delta q=-1$, and $+++\to ---$ gives
$\Delta q=-3$ with probability $\epsilon_{\downarrow}$. Therefore, for a given
configuration $\mathbf{s}$, the transition rates of these channels are
$W(q\to q+3|\mathbf{s})=\epsilon_{\uparrow}\Omega_0(\mathbf{s})$,
$W(q\to q+1|\mathbf{s})=\Omega_2(\mathbf{s})$,
$W(q\to q-1|\mathbf{s})=\Omega_1(\mathbf{s})$, and
$W(q\to q-3|\mathbf{s})=\epsilon_{\downarrow}\Omega_3(\mathbf{s})$. The mean
change of $q$ in configuration $\mathbf{s}$ is then
\begin{equation}
\langle \Delta q\rangle_{\mathbf{s}}
=
3\epsilon_{\uparrow}\Omega_0(\mathbf{s})
+
\Omega_2(\mathbf{s})
-
\Omega_1(\mathbf{s})
-
3\epsilon_{\downarrow}\Omega_3(\mathbf{s}).
\label{eq:mean_delta_q_config}
\end{equation}
The sign of $\langle \Delta q\rangle_{\mathbf{s}}$ indicates the local tendency
of the dynamics in configuration $\mathbf{s}$. The term $\Omega_2-\Omega_1$
represents the contribution of majority rule, while
$3\epsilon_{\uparrow}\Omega_0-3\epsilon_{\downarrow}\Omega_3$ represents the
contribution of collective reversal in unanimous triplets.

Equation~\eqref{eq:full_master_equation} is an exact description, but it involves
$2^N$ configurations and is therefore directly solvable only for small systems.
In a well-mixed population, triplet statistics can be written as functions of the
global density $c$, allowing the macroscopic dynamics to be obtained in closed
form. In contrast, on quenched networks, two configurations with the same value
of $c$ or $q$ may have different local arrangements. Consequently, local triplet
frequencies are not determined by the global variable alone, but also by the
network topology and by local correlations generated during the dynamics.

To build an aggregate description, the dynamics is projected onto the variable
$q=0,1,\ldots,N$, with $c=q/N$ and $m=2q/N-1$. This projection reduces the state
space from $2^N$ microscopic configurations to $N+1$ aggregate states. However,
because the variable $q$ alone does not close the dynamics on quenched networks,
we introduce the conditional triplet statistics
\begin{equation}
R_a(q;\epsilon)
=
\left\langle
\Omega_a(\mathbf{s})
\,\middle|\,
q(\mathbf{s})=q
\right\rangle_{\mathrm{st},\epsilon},
\qquad
a=0,1,2,3.
\label{eq:conditional_triplet_rate}
\end{equation}
Here, $\langle\cdots\rangle_{\mathrm{st},\epsilon}$ denotes an average over
stationary configurations at a given value of the collective reversal probability
$\epsilon$ and over quenched network realizations. In other words,
$R_a(q;\epsilon)$ is the average of $\Omega_a(\mathbf{s})$ over all stationary
configurations with the same number of positive spins. Since this average is
performed on quenched networks, $R_a(q;\epsilon)$ retains information about
degree heterogeneity, clustering, and local correlations that are absent in the
well-mixed approximation. This quantity is not obtained by directly fitting the
order-parameter curve, but from local triplet statistics measured from stationary
microscopic configurations. Thus, the projected master equation should be
understood as an effective Markovian description in the aggregate variable $q$,
with transition rates informed by local network statistics. The numerical
procedure for estimating $R_a(q;\epsilon)$ from stationary configurations is
briefly described in~\ref{app:mc_protocol}.

Using $R_a(q;\epsilon)$, the aggregate transition channels are given by
$T_{+3}(q;\epsilon)=\epsilon_{\uparrow}R_0(q;\epsilon)$,
$T_{+1}(q;\epsilon)=R_2(q;\epsilon)$,
$T_{-1}(q;\epsilon)=R_1(q;\epsilon)$, and
$T_{-3}(q;\epsilon)=\epsilon_{\downarrow}R_3(q;\epsilon)$. The mean change of
$q$ at the aggregate level is
\begin{equation}
\langle \Delta q\rangle_q
=
3\epsilon_{\uparrow}R_0(q;\epsilon)
+
R_2(q;\epsilon)
-
R_1(q;\epsilon)
-
3\epsilon_{\downarrow}R_3(q;\epsilon).
\label{eq:aggregate_drift_q}
\end{equation}
This equation shows that the aggregate drift is determined by the balance between
mixed triplets, which generate ordering through majority rule, and unanimous
triplets, which provide the channel for collective reversal.

For the symmetric collective-reversal case,
$\epsilon_{\uparrow}=\epsilon_{\downarrow}=\epsilon$, the projected master
equation for the aggregate probability $P(q,t)$ can be written as
\begin{align}
\frac{dP(q,t)}{dt}
=&\;
\epsilon R_0(q-3;\epsilon)P(q-3,t)
+
R_1(q+1;\epsilon)P(q+1,t)
\nonumber\\
&+
R_2(q-1;\epsilon)P(q-1,t)
+
\epsilon R_3(q+3;\epsilon)P(q+3,t)
\nonumber\\
&-
\left[
\epsilon R_0(q;\epsilon)
+
R_1(q;\epsilon)
+
R_2(q;\epsilon)
+
\epsilon R_3(q;\epsilon)
\right] P(q,t).
\label{eq:projected_master_equation}
\end{align}
The first four terms on the right-hand side represent probability currents
entering state $q$ from states $q-3$, $q+1$, $q-1$, and $q+3$, respectively.
The last term represents the probability current leaving state $q$. Thus,
Eq.~\eqref{eq:projected_master_equation} is a one-dimensional master equation in
the aggregate space $q$. Terms with indices outside the range $0\le q\le N$ are
discarded. The stationary distribution $P_{\mathrm{st}}(q)$ is obtained from
$dP(q,t)/dt=0$ with the normalization
$\sum_{q=0}^{N}P_{\mathrm{st}}(q)=1$.

Once the stationary distribution is obtained, the order parameter is calculated as
\begin{equation}
M(\epsilon)
=
\sum_{q=0}^{N}
\left|
\frac{2q}{N}-1
\right|
P_{\mathrm{st}}(q),
\label{eq:order_parameter_master}
\end{equation}
and is compared with the MC result,
$M_{\mathrm{MC}}(\epsilon)=\langle |m| \rangle$. The absolute value is required
because, in the symmetric case, the two ordered phases with positive and negative
magnetization are equivalent. Therefore, $M(\epsilon)$ measures the degree of
ordering without selecting either opinion direction.

With this formulation, MC simulations, the full master equation, and the
projected master equation can be compared using the same observable. For small
systems, the full master equation serves as an exact benchmark. For large systems,
the projected master equation provides an effective reduction that still retains
local information through $R_a(q;\epsilon)$.

\begin{figure}[t!]
    \centering
    \includegraphics[width=\linewidth]{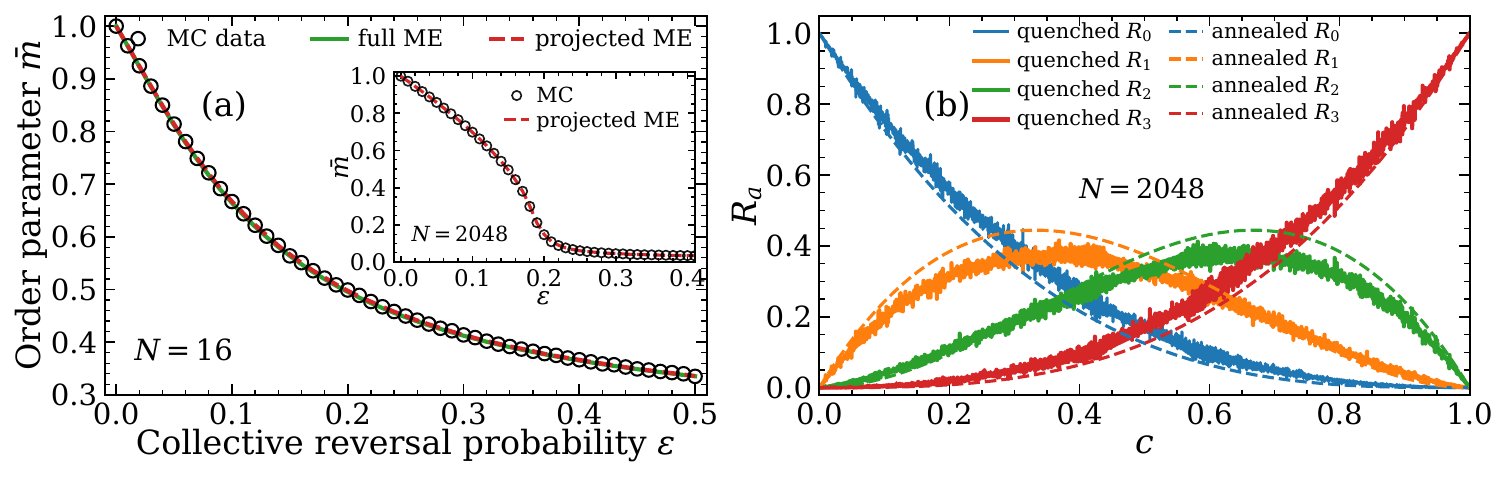}
    \caption{
    Validation of the master-equation description and local triplet statistics on
    a BA network with attachment parameter $m_{\mathrm{BA}}=4$,
    corresponding to $\langle k\rangle\simeq 8$. (a) Order parameter
    $\bar{m}=\langle |m| \rangle$ as a function of the symmetric collective
    reversal probability $\epsilon$ for $N=16$. The MC, full ME, and projected
    ME results are in very good agreement; the inset shows that the projected ME
    remains consistent with MC for $N=2048$. (b) Local triplet statistics
    $R_a$ as a function of $c=q/N$ for $N=2048$. Solid lines show the quenched
    results, while dashed lines show the annealed/mean-field prediction. The
    deviations between them highlight the role of topology and local correlations
    in triplet statistics.
    }
    \label{fig:master_validation}
\end{figure}

The comparison among these three approaches is shown in
Fig.~\ref{fig:master_validation}(a). For the small system $N=16$, the MC, full
ME, and projected ME results are in very good agreement. The inset in the same
panel shows that for $N=2048$, the projected ME still follows the MC results,
whereas the full ME is no longer computationally feasible because the
configuration space grows as $2^N$. Figure~\ref{fig:master_validation}(b)
shows the local triplet statistics $R_a$ as a function of the positive-opinion
density $c=q/N$. Solid lines show the results on the quenched network, whereas
dashed lines show the annealed/mean-field prediction. The deviations between the
two indicate that local triplet frequencies are determined not only by the global
density $c$, but also by network structure and local correlations.

The quantity $R_a(q;\epsilon)$ also provides a microscopic interpretation of the
shift in the critical point. In a well-mixed population, the probability of
obtaining a triplet with $a$ positive spins follows the binomial statistics
$R_a^{\mathrm{MF}}(c)=\binom{3}{a}c^a(1-c)^{3-a}$, and therefore depends only on
the global density $c$. On a quenched network, triplets are constrained by the
local neighborhood of the central node, so $R_a(q;\epsilon)$ generally deviates
from $R_a^{\mathrm{MF}}(c)$, as shown in Fig.~\ref{fig:master_validation}(b).
This deviation changes the balance between mixed and unanimous triplets. Mixed
triplets, namely $+--$ and $++-$, generate ordering through majority rule,
whereas unanimous triplets, namely $---$ and $+++$, provide the collective
reversal channels. Therefore, changes in the relative frequency of these two
classes of triplets can shift the critical point from the well-mixed value
$\epsilon_c^{\mathrm{MF}}=1/3$ (see~\ref{app:annealed_critical_point}).
Thus, the shift of $\epsilon_c$ on quenched networks can be understood as a
direct consequence of changes in local triplet statistics, rather than merely as a
finite-size effect.

\subsection{Monte Carlo results and finite-size scaling}
\label{subsec:fss}

To analyze the order--disorder transition of the local triplet model on quenched
networks, we first compute the order parameter, susceptibility, and Binder
cumulant from MC simulations. The order parameter is defined as the
mean absolute magnetization
\begin{equation}
M_N(\epsilon)
=
\left[
\left\langle |m| \right\rangle
\right]_{\mathrm{dis}},
\qquad
m=
\frac{1}{N}\sum_{i=1}^{N}s_i ,
\label{eq:fss_order_parameter}
\end{equation}
where $\langle \cdots \rangle$ denotes a time average or MC average in
the stationary state, while $[\cdots]_{\mathrm{dis}}$ denotes an average over
quenched network realizations. The use of $\langle |m|\rangle$ is necessary
because, in the symmetric reversal case, the two ordered states with positive and
negative magnetization are equivalent. Thus, the direct average $\langle m\rangle$
may approach zero even when the system remains in an ordered phase.

Fluctuations are measured through the susceptibility $\chi_N$, while the shape of
the magnetization distribution is characterized by the Binder cumulant $U_N$.
They are defined as~\cite{binder1981finite,privman1990finite,landau2021guide}
\begin{align}
\chi_N(\epsilon)
&=
N
\left[
\left\langle m^2 \right\rangle
-
\left\langle |m| \right\rangle^2
\right]_{\mathrm{dis}},
\\
U_N(\epsilon)
&=
\left[
1-
\frac{
\left\langle m^4 \right\rangle
}{
3\left\langle m^2 \right\rangle^2
}
\right]_{\mathrm{dis}} .
\label{eq:fss_binder}
\end{align}
The three quantities $M_N$, $\chi_N$, and $U_N$ are used together to identify
the transition region and estimate the critical point. Details of the MC
simulation protocol, including equilibration, number of realizations, parameter
grid selection, and stationary sampling procedure, are given in~\ref{app:mc_protocol}.

\begin{figure}[t!]
    \centering
    \includegraphics[width=\linewidth]{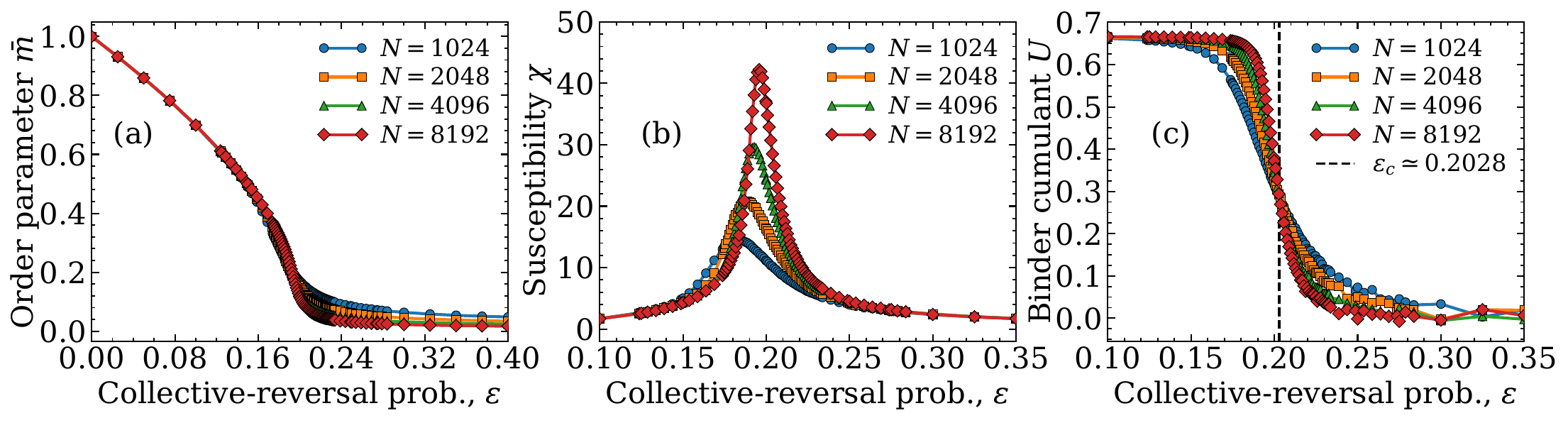}
\caption{
Monte Carlo simulation results on BA networks with attachment
parameter $m_{\mathrm{BA}}=4$, corresponding to $\langle k\rangle\simeq 8$.
(a) Order parameter $\bar{m}\equiv\langle |m| \rangle$, (b) susceptibility $\chi$,
and (c) Binder cumulant $U$ as functions of the symmetric collective reversal
probability $\epsilon$ for several system sizes $N$. The order parameter drops
sharply near the transition region, while the susceptibility peak becomes more
pronounced as $N$ increases. The crossing of the Binder cumulant curves gives an
estimate of the critical point \textcolor{black}{$\epsilon_c\simeq0.2028$}, indicated by the
vertical dashed line in panel (c). Around the critical region, each data point is
obtained from up to $10^5$ stationary samples and independent network
realizations.
}
    \label{fig:MC_BA_critical_point}
\end{figure}

Figure~\ref{fig:MC_BA_critical_point} shows the MC results for BA
networks with several system sizes. The BA network is generated with attachment
parameter $m_{\mathrm{BA}}=4$, giving an average degree close to
$\langle k\rangle\simeq 8$. As the symmetric collective reversal probability
$\epsilon$ increases, the order parameter $\langle |m|\rangle$ decreases from a
value close to one to a small value, indicating a change from the ordered phase to
the disordered phase. Near the transition region, the susceptibility $\chi$
develops a peak that becomes sharper for larger system sizes. Meanwhile, the
Binder cumulant curves for different $N$ intersect around the same value,
providing an initial estimate of the critical point. To improve the accuracy near
the transition, the control-parameter grid is refined to
$\Delta\epsilon=10^{-3}$ around the critical region. From the crossing of the
Binder cumulant curves, the best estimate for the BA network is
$\epsilon_c\simeq {\color{black}0.2028}$. This value is smaller than the critical
point of the corresponding well-mixed triplet model, $\epsilon_c=1/3$, which can
be obtained analytically or numerically.

After an initial critical point is obtained from the behavior of $\chi$ and $U$,
we perform a finite-size scaling analysis to estimate the critical exponents.
Because the system is defined on complex networks and has no characteristic
linear length as in Euclidean lattices, the scaling is performed directly with
the population size $N$. Thus, the relevant exponent is the correlation-volume
exponent $\bar{\nu}$, rather than the correlation-length exponent $\nu$.

For each network topology
$\mathcal{G}\in\{\mathrm{BA},\mathrm{ER},\mathrm{RR},\mathrm{WS}\}$, we assume
the finite-size scaling forms~\cite{privman1990finite,landau2021guide}
\begin{align}
M_N^{(\mathcal{G})}(\epsilon)
& =
N^{-\beta_{\mathcal{G}}/\bar{\nu}_{\mathcal{G}}}
f_m^{(\mathcal{G})}
\left[
\left(\epsilon-\epsilon_c^{(\mathcal{G})}\right)
N^{1/\bar{\nu}_{\mathcal{G}}}
\right],
\label{eq:fss_m} \\
\chi_N^{(\mathcal{G})}(\epsilon)
& =
N^{\gamma_{\mathcal{G}}/\bar{\nu}_{\mathcal{G}}}
f_\chi^{(\mathcal{G})}
\left[
\left(\epsilon-\epsilon_c^{(\mathcal{G})}\right)
N^{1/\bar{\nu}_{\mathcal{G}}}
\right],
\label{eq:fss_chi} \\
U_N^{(\mathcal{G})}(\epsilon)
& =
f_U^{(\mathcal{G})}
\left[
\left(\epsilon-\epsilon_c^{(\mathcal{G})}\right)
N^{1/\bar{\nu}_{\mathcal{G}}}
\right].
\label{eq:fss_binder_scaling}
\end{align}
Here, $\epsilon_c^{(\mathcal{G})}$ is the critical point for topology
$\mathcal{G}$, while $\beta_{\mathcal{G}}$, $\gamma_{\mathcal{G}}$, and
$\bar{\nu}_{\mathcal{G}}$ are the critical exponents for the order parameter,
susceptibility, and correlation volume, respectively.

At the critical point, Eqs.~\eqref{eq:fss_m} and \eqref{eq:fss_chi} yield
$M_N^{(\mathcal{G})}(\epsilon_c)\sim
N^{-\beta_{\mathcal{G}}/\bar{\nu}_{\mathcal{G}}}$ and
$\chi_{\max}^{(\mathcal{G})}(N)\sim
N^{\gamma_{\mathcal{G}}/\bar{\nu}_{\mathcal{G}}}$. Therefore, the ratios
$\beta_{\mathcal{G}}/\bar{\nu}_{\mathcal{G}}$ and
$\gamma_{\mathcal{G}}/\bar{\nu}_{\mathcal{G}}$ are obtained from the slopes of
the log-log plots of $M_N^{(\mathcal{G})}(\epsilon_c)$ and
$\chi_{\max}^{(\mathcal{G})}(N)$ versus $N$. The exponent
$1/\bar{\nu}_{\mathcal{G}}$ is estimated from the shift of the pseudo-critical
point. For finite systems, the susceptibility peak position
$\epsilon_c^{(\mathcal{G})}(N)$ shifts from the thermodynamic critical point
$\epsilon_c^{(\mathcal{G})}$ according to
$|\epsilon_c^{(\mathcal{G})}(N)-\epsilon_c^{(\mathcal{G})}|
\sim N^{-1/\bar{\nu}_{\mathcal{G}}}$. Thus, the slope of
$\ln |\epsilon_c^{(\mathcal{G})}(N)-\epsilon_c^{(\mathcal{G})}|$ versus
$\ln N$ gives $-1/\bar{\nu}_{\mathcal{G}}$. In the numerical analysis,
$\epsilon_c^{(\mathcal{G})}(N)$ is determined from the position of the
susceptibility peak using local interpolation around the critical region. The
obtained exponents are then used in the data-collapse procedure as a consistency
check.

\begin{figure}[t!]
    \centering
    \includegraphics[width=\linewidth]{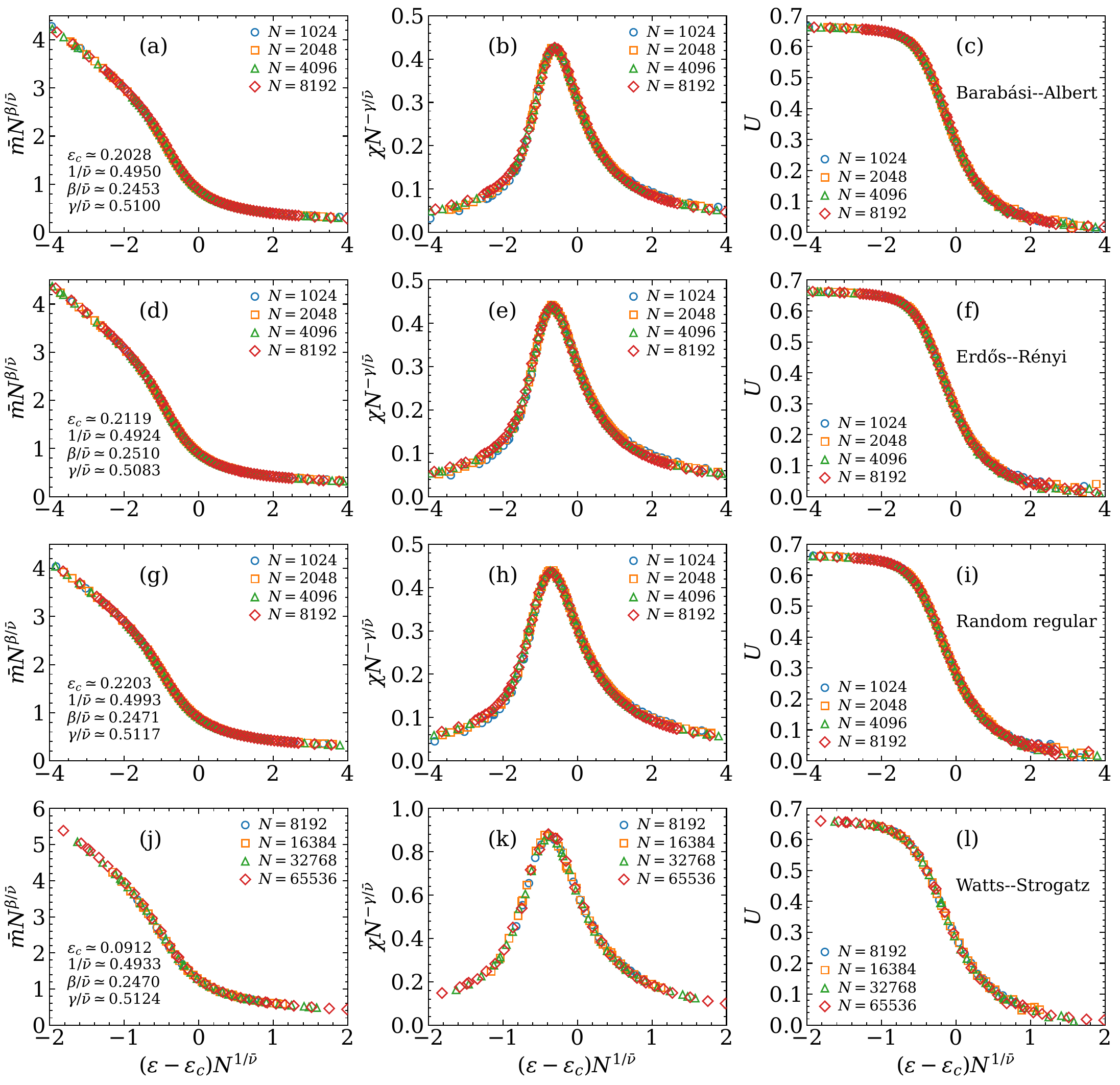}
    \caption{
    Finite-size scaling collapse for the triplet dynamics on four network
    types: Barabási--Albert (BA), Erdős--Rényi (ER), random regular (RR), and
    Watts--Strogatz (WS) networks. The first to fourth rows show the results
    for BA, ER, RR, and WS networks, respectively. The left column shows the
    scaled order parameter $\bar{m}N^{\beta/\bar{\nu}}$, the middle column shows
    the scaled susceptibility $\chi N^{-\gamma/\bar{\nu}}$, and the right column
    shows the Binder cumulant $U$ as a function of the scaling variable
    $(\epsilon-\epsilon_c)N^{1/\bar{\nu}}$.
    Different symbols denote different system sizes, {\color{black} from $N=1024$
    to $N = 8192$ for BA, ER, and RR networks, and from $N=8192$ to $N = 65536$
    for the WS network.
    }
    The values of $\epsilon_c$, $1/\bar{\nu}$, $\beta/\bar{\nu}$, and
    $\gamma/\bar{\nu}$ used to obtain the collapse are shown in the panels of the
    left column.
    }
    \label{fig:scaling_collapse_networks}
\end{figure}

In practice, the scaling variable is defined as
$x_{\mathcal{G}}=(\epsilon-\epsilon_c^{(\mathcal{G})})
N^{1/\bar{\nu}_{\mathcal{G}}}$. The values of $\epsilon_c^{(\mathcal{G})}$,
$1/\bar{\nu}_{\mathcal{G}}$, $\beta_{\mathcal{G}}/\bar{\nu}_{\mathcal{G}}$, and
$\gamma_{\mathcal{G}}/\bar{\nu}_{\mathcal{G}}$ are chosen so that curves for
different system sizes collapse when plotted as
$M_N^{(\mathcal{G})}N^{\beta_{\mathcal{G}}/\bar{\nu}_{\mathcal{G}}}$,
$\chi_N^{(\mathcal{G})}N^{-\gamma_{\mathcal{G}}/\bar{\nu}_{\mathcal{G}}}$, and
$U_N^{(\mathcal{G})}$ as functions of $x_{\mathcal{G}}$. A good collapse
indicates that data from different system sizes are consistent with a single
critical point and a single set of critical exponents for that topology. The
finite-size scaling results for BA, ER, RR, and WS networks are shown in
Fig.~\ref{fig:scaling_collapse_networks}. In the simulations, the network
parameters are chosen so that the average degree is comparable,
$\langle k\rangle\simeq 8$. Specifically, BA networks are generated with
attachment parameter $m_{\mathrm{BA}}=4$, ER networks with
$p_{\mathrm{ER}}=\langle k\rangle/(N-1)$ and $\langle k\rangle=8$, RR networks
with fixed degree $k_{\mathrm{RR}}=8$, and WS networks with $k_{\mathrm{WS}}=8$
and rewiring probability $\beta_{\mathrm{WS}}=0.10$. Thus, the comparison across
networks mainly highlights the effect of quenched topology, rather than a simple
difference in average degree.

\subsection{Topology dependence and drift-based interpretation}

Overall, the data for different system sizes show good collapse for each
topology, with the best estimates of the critical point and critical exponents
summarized in Table~\ref{tab:critical_exponents_networks}. As a reference, for
the well-mixed or mean-field system, the critical point of the homogeneous
triplet dynamics is $\epsilon_c^{\mathrm{MF}}=1/3$, with critical exponents
$\beta_{\mathrm{MF}}=1/2$, $\gamma_{\mathrm{MF}}=1$, and
$\bar{\nu}_{\mathrm{MF}}=2$. The results on quenched networks show that
$\epsilon_c$ is significantly shifted from the mean-field value. For BA, ER, and
RR networks, the critical point lies in the range
$\epsilon_c\simeq0.20$--$0.22$, below the well-mixed value
$\epsilon_c^{\mathrm{MF}}=1/3$. This means that a weaker symmetric collective
reversal is sufficient to destabilize the ordered phase on quenched networks. In
other words, restricting interactions to local neighbors makes global order less
stable than in the well-mixed system, where each agent can effectively interact
with the entire population.

From a social perspective, this result suggests that consensus formed through
fixed social contacts is more easily disrupted by external information than
consensus in a perfectly mixed population. In real social networks, individuals
usually interact only within limited environments, such as friends, communities,
or work groups. As a result, local agreement formed in one part of the network
may not be strong enough to sustain global order when all groups receive an
external perturbation that acts symmetrically on the two opinions. Thus, a fixed
social-relation structure can make collective opinion more vulnerable to external
perturbations, even when the perturbation does not explicitly favor either
opinion. This interpretation is consistent with recent models showing how network-induced correlations can destabilize global consensus and drive the system toward polarized states~\cite{baumann2020modeling}.

\begin{table}[bt!]
\centering
\caption{
Critical point and critical exponents of the triplet dynamics on different
quenched networks. The mean-field/well-mixed result is shown as a reference.
Numbers in parentheses indicate the estimated uncertainty in the last digit(s),
obtained from variations of the fitting window, system-size range, and collapse
quality. \textcolor{black}{The relative deviations are defined as
$\delta_x=|x-x_{\mathrm{MF}}|/x_{\mathrm{MF}}\times100\%$ for
$x=\bar{\nu},\beta,\gamma$.}
}
\vspace{0.4em}
\centering
\scriptsize
\setlength{\tabcolsep}{4pt}
\renewcommand{\arraystretch}{1.08}
\begin{tabular}{@{}lccccccc@{}}
\toprule
\multirow{2}{*}{Network}
& \multirow{2}{*}{$\epsilon_c$}
& \multicolumn{3}{c}{Critical exponents}
& \multicolumn{3}{c}{{\color{black}Relative deviation from MF (\%)}} \\
\cmidrule(lr){3-5}
\cmidrule(l){6-8}
& & $\bar{\nu}$ & $\beta$ & $\gamma$
& {\color{black}$\delta_{\bar{\nu}}$}
& {\color{black}$\delta_{\beta}$}
& {\color{black}$\delta_{\gamma}$} \\
\midrule
Mean-field / well-mixed
& $1/3$
& $2$
& $1/2$
& $1$
& {\color{black}$0$}
& {\color{black}$0$}
& {\color{black}$0$} \\
Barabási--Albert
& {\color{black}$0.2028(3)$}
& {\color{black}$2.02(4)$}
& {\color{black}$0.50(1)$}
& {\color{black}$1.03(2)$}
& {\color{black}$1.0$}
& {\color{black}$0.0$}
& {\color{black}$3.0$} \\
Erdős--Rényi
& {\color{black}$0.2119(2)$}
& {\color{black}$2.03(4)$}
& {\color{black}$0.51(3)$}
& {\color{black}$1.03(6)$}
& {\color{black}$1.5$}
& {\color{black}$2.0$}
& {\color{black}$3.0$} \\
Random regular
& {\color{black}$0.2203(5)$}
& {\color{black}$2.00(6)$}
& {\color{black}$0.49(3)$}
& {\color{black}$1.02(6)$}
& {\color{black}$0.0$}
& {\color{black}$2.0$}
& {\color{black}$2.0$} \\
Watts--Strogatz
& {\color{black}$0.0912(4)$}
& {\color{black}$2.03(5)$}
& {\color{black}$0.50(3)$}
& {\color{black}$1.04(2)$}
& {\color{black}$1.5$}
& {\color{black}$0.0$}
& {\color{black}$4.0$} \\
\bottomrule
\end{tabular}
\label{tab:critical_exponents_networks}
\end{table}

Interestingly, although the critical point is shifted from the well-mixed value,
the critical exponents for {\color{black}all network classes considered} remain
close to the mean-field values. For BA, ER, and RR networks, the exponent
$\beta$ lies around {\color{black}$0.49$--$0.51$}, very close to
$\beta_{\mathrm{MF}}=1/2$, while $\gamma$ lies around
{\color{black}$1.02$--$1.03$}, close to $\gamma_{\mathrm{MF}}=1$. The
correlation-volume exponent $\bar{\nu}$ is also close to
$\bar{\nu}_{\mathrm{MF}}=2$, with values around {\color{black}$2.00$--$2.03$}.
Thus, these networks exhibit mean-field-like critical behavior: the quenched
topology shifts the transition point without substantially altering the critical
exponents.

{\color{black}The WS network shows the strongest shift of the critical point, with
$\epsilon_c=0.0912(4)$, much lower than the values obtained for BA, ER, and RR
networks.} However, its estimated exponents,
{\color{black}$\bar{\nu}=2.03(5)$, $\beta=0.50(3)$, and $\gamma=1.04(2)$},
also remain close to the mean-field values within the numerical uncertainties.
{\color{black}Therefore, the main effect of the WS topology is not a clear change of
universality class, but a strong downward shift of the transition point.} This
shift is naturally associated with the stronger clustering and persistent local
correlations of the WS network, which modify the balance between mixed and
unanimous local triplets and make the ordered phase less stable against
collective reversal.

To further examine this interpretation, we estimate the critical point on WS
networks for several values of the rewiring probability $\beta_{\mathrm{WS}}$
using a drift-based approach built from local triplet statistics. This approach is
complementary to finite-size scaling: finite-size scaling determines
$\epsilon_c$ from the collective behavior of systems with several sizes $N$,
whereas the local drift approach estimates the stability change of the fixed
point $c=1/2$ from local triplet statistics at a fixed system size.

In the symmetric collective-reversal case, the aggregate drift can be written as
$F(c;\epsilon)=3\epsilon[R_0(c;\epsilon)-R_3(c;\epsilon)]
+R_2(c;\epsilon)-R_1(c;\epsilon)$. Due to up--down symmetry, the disordered
state $c=1/2$ is a fixed point. The local stability of this fixed point is
determined by
$\lambda(\epsilon)=\left.\partial F(c;\epsilon)/\partial c\right|_{c=1/2}$,
and the local critical point is estimated from the condition
$\lambda(\epsilon_c)=0$. Equivalently, by defining
$R_a'(1/2;\epsilon)=\left.\partial R_a(c;\epsilon)/\partial c\right|_{c=1/2}$,
the local critical point can be expressed as the implicit relation
\begin{equation}
\epsilon_c^{\mathrm{local}}
=
-
\frac{
R_2'(1/2;\epsilon_c)-R_1'(1/2;\epsilon_c)
}{
3\left[
R_0'(1/2;\epsilon_c)-R_3'(1/2;\epsilon_c)
\right]
}.
\label{eq:local_critical_point}
\end{equation}
In the numerical calculation, $R_a'(1/2;\epsilon)$ is obtained from a local fit
of $R_a(c;\epsilon)$ around $c=1/2$ using the fitting window $w=0.06$. Since
$R_a(c;\epsilon)$ is measured from stationary configurations at a given value of
$\epsilon$, Eq.~\eqref{eq:local_critical_point} should be interpreted as a local
stability estimate based on triplet statistics, rather than as a direct substitute
for the finite-size scaling estimate.

\begin{figure}[bt!]
    \centering
    \includegraphics[width=\linewidth]{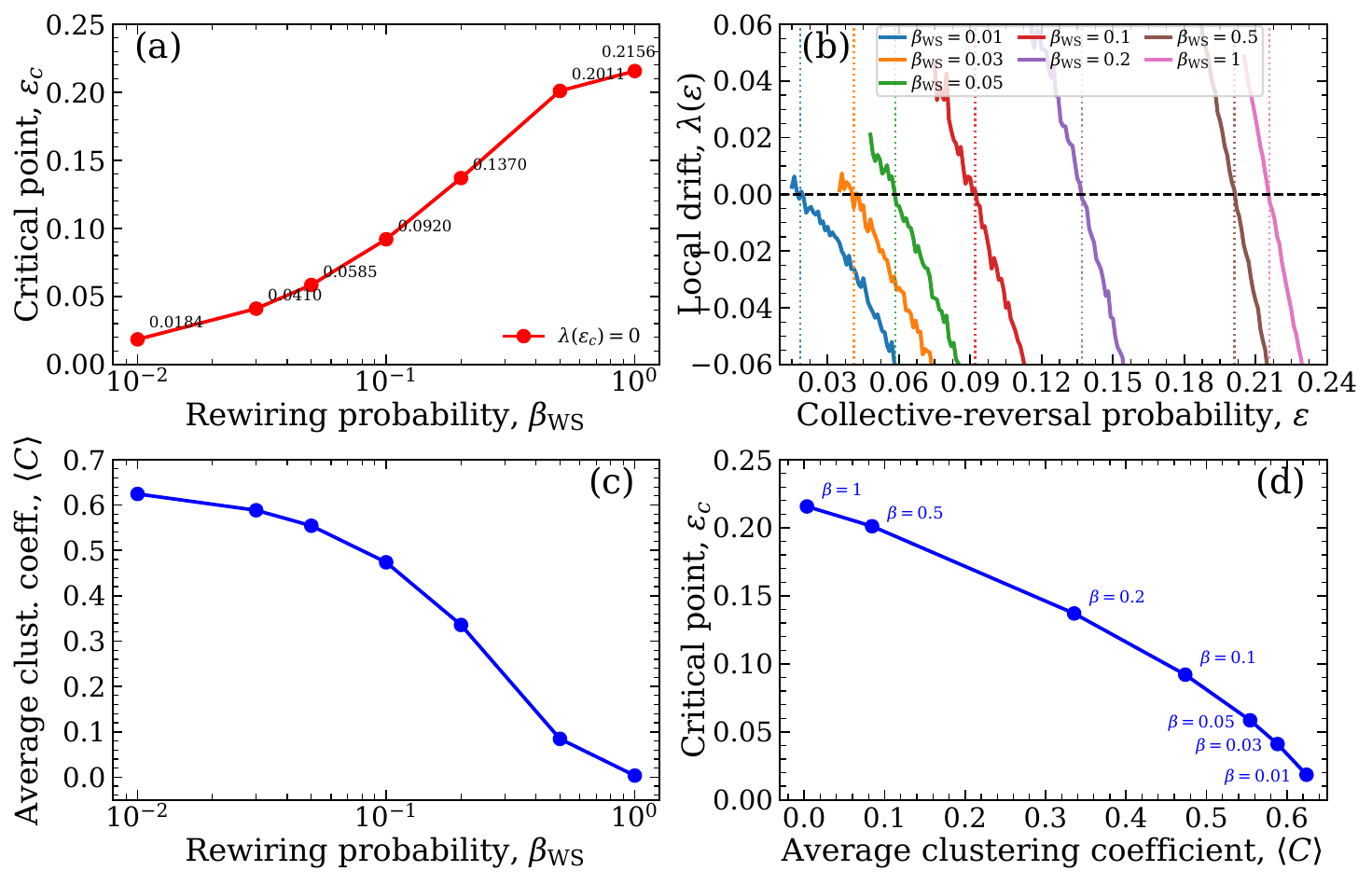}
    \caption{
{\color{black}
Rewiring and clustering effects on the critical point of WS networks.
(a) Critical point $\epsilon_c$ as a function of the rewiring probability
$\beta_{\mathrm{WS}}$ for $N=2048$ and $k_{\mathrm{WS}}=8$, obtained from
$\lambda(\epsilon_c)=0$. (b) Local drift $\lambda(\epsilon)$ for different
$\beta_{\mathrm{WS}}$, with vertical dotted lines indicating the corresponding
$\epsilon_c$. (c) Average clustering coefficient $\langle C\rangle$ as a
function of $\beta_{\mathrm{WS}}$. (d) Critical point $\epsilon_c$ as a function
of $\langle C\rangle$, with labels indicating the corresponding
$\beta_{\mathrm{WS}}$. For $\beta_{\mathrm{WS}}=0.10$,
$\epsilon_c^{\mathrm{local}}\simeq0.0920$, in good agreement with
$\epsilon_c^{\mathrm{FSS}}=0.0912(4)$. 
The results show that rewiring reduces clustering and increases the critical
point, linking the low $\epsilon_c$ of WS networks to strong clustering and local
correlations.}
}
    \label{fig:estimasi_WS}
\end{figure}

By keeping $k_{\mathrm{WS}}=8$ and varying $\beta_{\mathrm{WS}}$, this analysis
separates the effect of average connectivity from that of local correlations.
{\color{black}
The local-drift estimates in Fig.~\ref{fig:estimasi_WS}(a) show that the critical
point $\epsilon_c$ increases monotonically as $\beta_{\mathrm{WS}}$ is increased.
This means that the ordered phase becomes more stable as the WS network becomes
more random. The corresponding local-drift curves in
Fig.~\ref{fig:estimasi_WS}(b) show the change of stability through the crossing
condition $\lambda(\epsilon_c)=0$. To make the connection with clustering more quantitative, we also measured the
average clustering coefficient $\langle C\rangle$ for the same WS networks. As
shown in Fig.~\ref{fig:estimasi_WS}(c), $\langle C\rangle$ decreases
monotonically with increasing $\beta_{\mathrm{WS}}$, confirming that rewiring
systematically reduces local clustering. Figure~\ref{fig:estimasi_WS}(d) further
shows that $\epsilon_c$ decreases as $\langle C\rangle$ increases. Thus, the low
critical point of WS networks can be quantitatively associated with their large
clustering coefficient and persistent local correlations. In this sense,
clustering weakens the stability of the ordered phase against collective reversal,
whereas rewiring reduces clustering and increases the critical point.
}

For the WS network used in the finite-size scaling analysis,
$\beta_{\mathrm{WS}}=0.10$, Table~\ref{tab:critical_exponents_networks} gives
$\epsilon_c^{\mathrm{FSS}}={\color{black}0.0912(4)}$, whereas the local drift
approach in Eq.~\eqref{eq:local_critical_point} gives
$\epsilon_c^{\mathrm{local}}\simeq0.0920$, as shown in
Fig.~\ref{fig:estimasi_WS}{\color{black}(a,b)}. The absolute difference between
the two estimates is
$|\epsilon_c^{\mathrm{local}}-\epsilon_c^{\mathrm{FSS}}|\simeq
{\color{black}0.0008}$, corresponding to a relative difference of about
{\color{black}$0.9\%$}. {\color{black}This close agreement} is reasonable because
the two methods probe different aspects of the critical dynamics: finite-size
scaling uses data from several system sizes and a collapse procedure, whereas the
local drift estimate uses derivatives of triplet statistics around $c=1/2$ for a
fixed system size. Therefore, {\color{black}agreement at the sub-percent level}
indicates that the local drift approach {\color{black}provides a reliable}
microscopic diagnostic of the critical-point shift, especially for explaining the
effects of rewiring, clustering, and local correlations in WS networks.

As a consistency check, we also examine the hyperscaling relation in volume form,
$\bar{\nu}\simeq 2\beta+\gamma$. Using the central estimates in
Table~\ref{tab:critical_exponents_networks}, we find that this relation is
{\color{black}satisfied well for all network classes}. For the BA network, for
example, $\bar{\nu}={\color{black}2.02}$, while
$2\beta+\gamma\simeq{\color{black}2.03}$, giving a relative deviation of about
{\color{black}$0.5\%$}. For the ER network,
$\bar{\nu}={\color{black}2.03}$ and
$2\beta+\gamma\simeq{\color{black}2.05}$, corresponding to a deviation of about
{\color{black}$1.0\%$}. For the RR network,
$\bar{\nu}={\color{black}2.00}$ and
$2\beta+\gamma\simeq{\color{black}2.00}$, so the agreement is
{\color{black}essentially exact within the numerical precision}. For the WS
network, $\bar{\nu}={\color{black}2.03}$ and
$2\beta+\gamma\simeq{\color{black}2.04}$, giving a relative deviation of about
{\color{black}$0.5\%$}. These small deviations indicate that the estimated critical exponents are
consistent with the volume-form hyperscaling relation within the numerical
uncertainties. Therefore, the finite-size scaling analysis supports a
mean-field-like critical regime for {\color{black}all network classes considered}.
In particular, the WS network {\color{black}satisfies the hyperscaling relation as
well as the other network classes}, indicating that its main distinction is the
{\color{black}much lower critical point rather than a clear departure from
mean-field-like critical behavior}.

This comparison shows that the finite-size scaling behavior of triplet dynamics
on quenched networks is more topology-dependent than in the well-mixed case. For
random networks with comparable average degree, such as BA, ER, and RR, the
critical behavior remains in a mean-field-like regime. However, when the network
has stronger {\color{black}clustering and} local correlations, as in the WS
network, {\color{black}the critical point is shifted much more strongly, even
though the effective exponents remain close to the mean-field values.} Therefore,
in addition to the symmetry of the opinion variable and the size of the state
space, network topology also plays an important role in {\color{black}setting the
transition point and shaping the finite-size critical behavior of the model.}

\section{\label{sec:summary}Summary and Conclusion}

In this study, we investigated an opinion-dynamics model based on triplet
majority rule with collective reversal on quenched networks. In contrast to the
well-mixed model, interactions in the present model are constrained by a fixed
network structure, so opinion updates take place on local triplets consisting of
one central agent and two of its neighbors. Majority rule acts on mixed triplets,
whereas collective reversal acts only on triplets that have reached a fully
unanimous state. Thus, the model clearly separates two main mechanisms: local
conformity through majority rule and external perturbations that can reverse
local agreement.

To describe the aggregate dynamics on quenched networks, we used a projected
master-equation formulation based on the conditional local-triplet statistics
$R_a(q;\epsilon)$. This approach retains information about local network
structure and correlations that are not captured by a homogeneous mean-field
approximation. The validation results show that the projected master equation
reproduces the MC results and, for small systems, is consistent with the
full master equation. Therefore, local triplet statistics provide an effective
description that connects microscopic dynamics on fixed networks with the
macroscopic behavior of the system. However, since $R_a(q;\epsilon)$ is obtained
from stationary microscopic configurations, this description should be understood
as an effective Markovian reduction rather than a fully closed analytical theory.

Monte Carlo simulations and finite-size scaling analysis reveal a
topology-dependent order--disorder transition. Compared with the well-mixed case,
where the critical point is $\epsilon_c^{\mathrm{MF}}=1/3$, the critical point on
quenched networks is shifted to lower values. For BA, ER, and RR networks, we
find
{\color{black}
$\epsilon_c=0.2028(3)$, $\epsilon_c=0.2119(2)$, and
$\epsilon_c=0.2203(5)$,
}
respectively, with critical exponents that remain close to the mean-field
values. This indicates that, for these three networks, topology mainly shifts the
location of the transition without drastically changing the class of critical
behavior.

{\color{black}
The WS network shows the strongest shift of the critical point, with
$\epsilon_c=0.0912(4)$, which is much lower than the values obtained for BA, ER,
and RR networks. However, the corresponding critical exponents remain close to
the mean-field values within the numerical uncertainties. This indicates that
clustering, local correlations, and the small-world structure mainly weaken the
stability of the ordered phase by shifting the transition point, rather than
producing clear evidence of a distinct universality class.
}
This interpretation is also supported by the local drift analysis based on
triplet statistics. For the WS network with $\beta_{\mathrm{WS}}=0.10$, the local
drift approach gives $\epsilon_c^{\mathrm{local}}\simeq0.0920$,
{\color{black}
in good agreement with the finite-size scaling estimate
$\epsilon_c=0.0912(4)$, with a relative difference of about $0.9\%$.
}
This agreement shows that the local drift based on $R_a(q;\epsilon)$ provides a
useful microscopic diagnostic for understanding the shift of the critical point
caused by network structure.

These results show that quenched network topology
{\color{black}
primarily shifts the location of the transition, while the estimated critical
exponents remain close to the mean-field values for all network classes
considered.
}
Local triplet statistics provide a mechanism for explaining how network
structure changes the balance between ordering by majority rule and collective
reversal in unanimous states.
{\color{black}
The WS topology provides the clearest example of this effect: its stronger
clustering and local correlations substantially lower the critical point, even
though its finite-size scaling exponents remain mean-field-like.
}
This study offers a simple framework for understanding opinion dynamics based on
small-group interactions on fixed social structures. Natural extensions of this
work include asymmetric collective reversal, weighted or directed networks, and
networks with stronger community structure.

\section*{Acknowledgments}
\textbf{R.~Muslim} acknowledges support from the YST Program of the Asia Pacific Center for Theoretical Physics (APCTP), funded by the Science and Technology Promotion Fund and the Lottery Fund of the Korean Government, and from the Management Talent Program of the National Research and Innovation Agency of Indonesia (BRIN).

\section*{ORCID}
\noindent Roni Muslim - \url{https://orcid.org/0000-0001-6925-5923}

\appendix
\section{Monte Carlo simulation protocol}
\label{app:mc_protocol}

Monte Carlo simulations are performed on quenched networks using the local
triplet update rule described in Sec.~\ref{sec:model}. One elementary update
starts by selecting a central node $r$ at random; if $k_r<2$, the selection is
repeated until a node with at least two neighbors is obtained. Two distinct
neighbors $j,\ell\in\partial r$ are then selected uniformly at random to form the
triplet $\tau=(r,j,\ell)$. Mixed triplets are updated according to the majority
rule, whereas unanimous triplets may undergo collective reversal. In the main
simulations, we consider the symmetric case
$\epsilon_{\uparrow}=\epsilon_{\downarrow}=\epsilon$. One Monte Carlo step
(MCS) is defined as $N$ elementary updates. Four types of networks are used:
BA, ER, RR, and
WS, with parameters chosen so that
$\langle k\rangle\simeq8$: $m_{\mathrm{BA}}=4$, $\langle k\rangle=8$ for ER,
$k_{\mathrm{RR}}=8$, and $k_{\mathrm{WS}}=8$. For ER and WS networks, only
connected realizations with minimum degree $k_i\geq2$ are used.

For the finite-size scaling analysis, the system sizes are
{\color{black}
$N=1024,2048,4096$, and $8192$ for BA, ER, and RR networks, while
$N=8192,16384,32768$, and $65536$ for the WS network.
}
The control parameter $\epsilon$ is chosen using an adaptive grid: a coarse grid
over the full range $0\leq\epsilon\leq0.5$, a medium grid around the critical
region, and a fine grid with spacing $\Delta\epsilon=10^{-3}$ near the critical
point. The number of realizations is also adaptive, with 100, 250, and 500
realizations for the coarse, medium, and fine grids, respectively. Each run
starts from a random initial condition with probability $p_{\mathrm{init}}=0.5$
for positive spins. Equilibration is performed adaptively with a minimum of
300 MCS and a maximum of 6000 MCS. During equilibration, block averages of $|m|$
are computed every 100 MCS; the system is considered stationary when the last
three block averages differ by less than $2.0\times10^{-3}$. After
equilibration, measurements are performed for 4000 MCS, and configurations are
sampled every 20 MCS. From the stationary magnetization series, $M_N$, $\chi_N$,
and $U_N$ are computed as defined in Sec.~\ref{sec:results}.

For the validation of the projected master equation, the conditional triplet
statistics $R_a(q;\epsilon)$ are estimated from stationary MC
configurations. For each measured configuration, the value of $q$ is recorded,
and local triplets are sampled using the same selection rule as in the dynamics.
The fraction of triplets with $a=0,1,2,3$ positive spins is then averaged over
all configurations with the same value of $q$ and over network realizations. In
the comparison between MC and the projected ME for $N=2048$, we use 500
realizations for each value of $\epsilon$, 4000 MCS for measurement, sampling
every 20 MCS, and 2000 sampled triplets per configuration. Thus, each value of
$\epsilon$ provides $500\times(4000/20)=10^5$ stationary configurations for
estimating the triplet statistics. For small systems, the MC and
projected ME results are also compared with the full master equation.

For the drift-based estimate of the critical point on WS networks, we use
$N=2048$, $k_{\mathrm{WS}}=8$, and several values of the rewiring probability
$\beta_{\mathrm{WS}}$. For each $\beta_{\mathrm{WS}}$, the parameter $\epsilon$
is taken on a local grid with spacing $\Delta\epsilon=10^{-3}$ around the
critical region. The triplet statistics $R_a(c;\epsilon)$ are computed from
stationary configurations, and the derivatives $R_a'(1/2;\epsilon)$ are obtained
from a weighted linear fit in the window $|c-1/2|\leq0.06$. The local drift is
defined as
$F(c;\epsilon)=3\epsilon[R_0(c;\epsilon)-R_3(c;\epsilon)]
+R_2(c;\epsilon)-R_1(c;\epsilon)$, while the stability of the disordered state is
determined by
$\lambda(\epsilon)=\left.\partial F/\partial c\right|_{c=1/2}$. The local
critical point is obtained from the condition $\lambda(\epsilon_c)=0$, using
linear interpolation between two consecutive values of $\epsilon$ where
$\lambda$ changes sign.

{\color{black}
For the clustering analysis in Fig.~\ref{fig:estimasi_WS}(c,d), the average
clustering coefficient $\langle C\rangle$ is computed for WS networks with the
same parameters, $N=2048$ and $k_{\mathrm{WS}}=8$, over the same set of rewiring
probabilities $\beta_{\mathrm{WS}}$. For each network realization, the local
clustering coefficient of node $i$ is calculated as
$C_i=2e_i/[k_i(k_i-1)]$ for $k_i\geq2$, where $e_i$ is the number of links among
the neighbors of node $i$. The network clustering coefficient is then obtained as
$C=N^{-1}\sum_i C_i$, and $\langle C\rangle$ is averaged over independent WS
network realizations. In the present calculation, this average is computed over
$1000$ independent network realizations for each value of $\beta_{\mathrm{WS}}$.
}

\section{Critical point in the annealed limit}
\label{app:annealed_critical_point}

In the annealed or well-mixed limit, triplets are selected randomly from the
entire population. If $c$ is the density of agents with opinion $+1$, the
probability of selecting a triplet with $a$ positive spins follows the binomial
distribution
$R_a^{\mathrm{MF}}(c)=\binom{3}{a}c^a(1-c)^{3-a}$, with $a=0,1,2,3$. For the
symmetric collective-reversal case,
$\epsilon_{\uparrow}=\epsilon_{\downarrow}=\epsilon$, the mean-field drift is
$F_{\mathrm{MF}}(c;\epsilon)
=3\epsilon[R_0^{\mathrm{MF}}(c)-R_3^{\mathrm{MF}}(c)]
+R_2^{\mathrm{MF}}(c)-R_1^{\mathrm{MF}}(c)$. Substituting the binomial form gives
\begin{equation}
F_{\mathrm{MF}}(c;\epsilon)
=
3(2c-1)
\left[
c(1-c)-\epsilon(1-c+c^2)
\right].
\label{eq:FMF_factorized}
\end{equation}
Due to up--down symmetry, $c=1/2$ is always a fixed point. Its local stability is
determined by
$\lambda_{\mathrm{MF}}(\epsilon)
=\left.\partial F_{\mathrm{MF}}/\partial c\right|_{c=1/2}
=\frac{3}{2}(1-3\epsilon)$. The critical point is obtained from
$\lambda_{\mathrm{MF}}(\epsilon_c)=0$, yielding
$\epsilon_c^{\mathrm{MF}}=1/3$.

\bibliographystyle{ws-ijmpc}
\bibliography{ws-sample}

\end{document}